\documentclass[prl,reprint,amsmath,amssymb,twocolumn,showpacs]{revtex4}
\usepackage{ulem}
\usepackage{color}

\usepackage{verbatim}
\usepackage{graphicx}
\usepackage{hyperref}
\hypersetup{
    colorlinks=true,
    urlcolor=magenta,
}
\def\pmb#1{\setbox0=\hbox{#1}
\kern-.025em\copy0\kern-\wd0 \kern-.05em\copy0\kern-\wd0
\kern-.025em\raise.0433em\box0}

\newcommand{\beq}{\begin{equation}}
\newcommand{\eeq}{\end{equation}}
\newcommand{\ba}{\begin{eqnarray}}
\newcommand{\ea}{\end{eqnarray}}
\font\myfont=cmr12 at 15pt

\newif\iffigures

\begin{document}

\title[]{\hspace{0.70cm} Observations of symmetry induced topological mode steering in a \newline reconfigurable elastic plate
}






\author{K. Tang$^1$, M. Makwana$^{2, 3}$, R.~V. Craster$^{2,4}$ \& P. Sebbah$^{1,5}$}
\affiliation{$^1$ Department of Physics, The Jack and Pearl Resnick Institute for Advanced Technology,
Bar Ilan University, Ramat Gan 5290002, Israel}
\affiliation{$^2$ Department of Mathematics, Imperial College London, London SW7 2AZ, UK}
\affiliation{$^3$ Multiwave Technologies AG, 3 Chemin du Pr\^{e} Fleuri, 1228, Geneva, Switzerland}
\affiliation{$^4$ Department of Mechanical Engineering, Imperial College London, London SW7 2AZ, UK}
\affiliation{$^5$ Institut Langevin, ESPCI ParisTech CNRS UMR7587, 1 rue Jussieu, 75238 Paris cedex 05, France}

\begin{abstract}
We experimentally investigate the valley-Hall effect for interfacial edge states, highlighting the importance of the modal patterns, between geometrically distinct regions within a structured elastic plate. These experiments, for vibration, are at a scale where detailed measurements are taken throughout the system and not just at the input/output ports; this exposes the coupling between geometrically distinct modes that   
 underlie the  
 differences between wave transport around gentle and sharp bends.
 

\end{abstract}
\maketitle

A dominant theme in  wave physics is the profound importance of material, or geometrical, structurations on the global behaviour of waves through a medium. Periodic media have been particularly influential, with  Bragg scattering  having created the fields of photonic crystals in optics \cite{joannopoulos_photonic_2008,zolla_foundations_2005}, and their acoustic counterparts in phononic systems \cite{laude_phononic_2015}. Demands for systems 
 enabling the control, of low-frequency vibration in sub-wavelength devices motivated additional resonant elements leading to the field of metamaterials \cite{liu_locally_2000,pendry_science_2006}. A further natural progression has drawn upon recent  advances in topological insulators \cite{hasan2010colloquium} and recently topological concepts have emerged as a new degree of freedom to control the flux of energy across diverse fields, which originated in condensed matter physics and then expanded to Newtonian wave systems \cite{Kane2014,Vitelli12266,Mousavi2015,Khanikaev2015,Fleury2016,zhang_topological_2018,ye_observation_2017, zhang_manipulation_2018, yan_on-chip_2018, xiaoxiao_direct_2017, shalaev_experimental_2017, ma_all-si_2016, lu_observation_2016, liu_experimental_2018, kang_pseudo-spinvalley_2018, jung_active_2018, gao_valley_2017, cheng_robust_2016, chen_tunable_2018, chen2019valley}. 

 Applying these ideas, from topological insulators, to elastic vibration are naturally attractive, although such systems lack the many degrees of freedom available to quantum systems  
 from where these ideas originate. Active topological systems require the absence of time-reversal symmetry (TRS) and in elastic systems this is achieved via the introduction of gyroscopes \cite{carta_2015,wang_2015}. Breaking TRS is harder for bosonic systems, such as those emerging in elastic waves without complex experimental set-ups and, in practical terms one is naturally drawn to considering passive valley-Hall systems that simply break spatial symmetries to induce quasi-topological modes. Despite these modes not being as robust, as their TRS breaking counterparts, they offer a pragmatic route to leverage a few of the important protective properties associated with the TRS broken systems. For valley-Hall systems the gapping of Dirac cones leaves behind two pronounced valleys, upon which a nontrivial Berry phase is defined, that contribute to the quasi-topological modes; this has led to a research area broadly referred to as valleytronics \cite{schaibley2016valleytronics}. 


We investigate elastic plate models \cite{landau_theory_1970,graff_wave_1975} that are highly effective at  providing reliable predictions of many wave phenomena for
 structured plates \cite{lefebvre_unveiling_2017}, platonic models utilising Dirac
 cones in the style of graphene 
 \cite{torrent_elastic_2013} and valley-Hall states \cite{pal_edge_2017,makwana_geometry_2018};
  the experiments of \cite{lefebvre_unveiling_2017} show remarkable agreement between theory and experiment even outside the conventional assumptions of validity of the model. The plate model also acts
 in practical terms as motivation for seismic metamaterials
 \cite{brule_experiments_2014,colombi_seismic_2016}. The model is also popular at the nano-scale, with  \cite{yan_on-chip_2018} recently considering high frequency ultrasonic applications, at MHz frequencies, and 100 micron thick, elastic plates.
 
 

To take full advantage of valley-Hall states, as new potential carriers of information, it is imperative to understand the coupling between the geometrically distinct modes. 
In this Letter we highlight the importance of these edge-state modes for valley-Hall systems and provide macroscopic 
 experimental results showing distinctive modal patterns that have, thus far, been absent in the majority of earlier studies \cite{zhang_topological_2018,ye_observation_2017, zhang_manipulation_2018, yan_on-chip_2018, xiaoxiao_direct_2017, shalaev_experimental_2017, ma_all-si_2016, lu_observation_2016, liu_experimental_2018, kang_pseudo-spinvalley_2018, jung_active_2018, gao_valley_2017, cheng_robust_2016, chen_tunable_2018, chen2019valley} and demonstrate their role in interpreting relative interface orientations.  The wealth of experimental demonstrations using these valley-Hall states has primarily focussed on electromagnetism \cite{zhang_manipulation_2018, shalaev_experimental_2017, ma_all-si_2016, kang_pseudo-spinvalley_2018, gao_valley_2017, cheng_robust_2016, chen_tunable_2018, chen2019valley} and acoustics \cite{zhang_topological_2018,ye_observation_2017, lu_observation_2016} at scales where mode details,  particularly their mode shapes, have been indistinct; the subtleties of different modes on different interfaces \cite{makwana_geometry_2018}, or even that there are different modes, 
 has often been overlooked. And yet an appreciation of the geometrically distinct modes, and the coupling between them, is needed in order to utilise them in more complex domains \cite{makwana_design_2018}.
The majority of experimental valleytronic papers, to name but a few \cite{zhang_topological_2018, zhang_manipulation_2018, yan_on-chip_2018, xiaoxiao_direct_2017, shalaev_experimental_2017, lu_observation_2016, liu_experimental_2018, jung_active_2018, gao_valley_2017, chen_tunable_2018, chen2019valley,jin2018}, use a  Z-shaped interface to demonstrate robustness of quasi-topological modes. There are two basic bend-types, sharp and gentle bends; these are 
 defined by the angle through which  energy is redirected, i.e. $2\pi/3$ and $ \pi/3$ respectively. We examine a gentle bend, in addition to the sharp bend (present within the Z-shaped  interface) and draw attention to the modal conversion that occurs for the former, but not the latter. This crucial property, that underlies the construction of topological networks, is demonstrated experimentally for a structured elastic plate which has the advantage of being macro-scale and hence these details are experimentally accessible.


 We also address the  pressing question of how effective the emerging topological concepts from valleytronics are for vibration systems in practical terms; particularly at centimetric scales and kHz frequencies
  relevant to engineering structures. 
  We perform careful experiments and cross-validate with recent theoretical valleytronic designs for elastic plates  \cite{makwana_geometry_2018}; these involve an infinite hexagonal structure, with each cell containing idealised point masses arranged to lie at the vertices of triangles. It is also interesting to observe the similarities and, more importantly, the crucial differences between the experiments and the idealised theoretical results.
  
  

 



\begin{figure}[h!]
\centering 
  \includegraphics[width = 8cm]{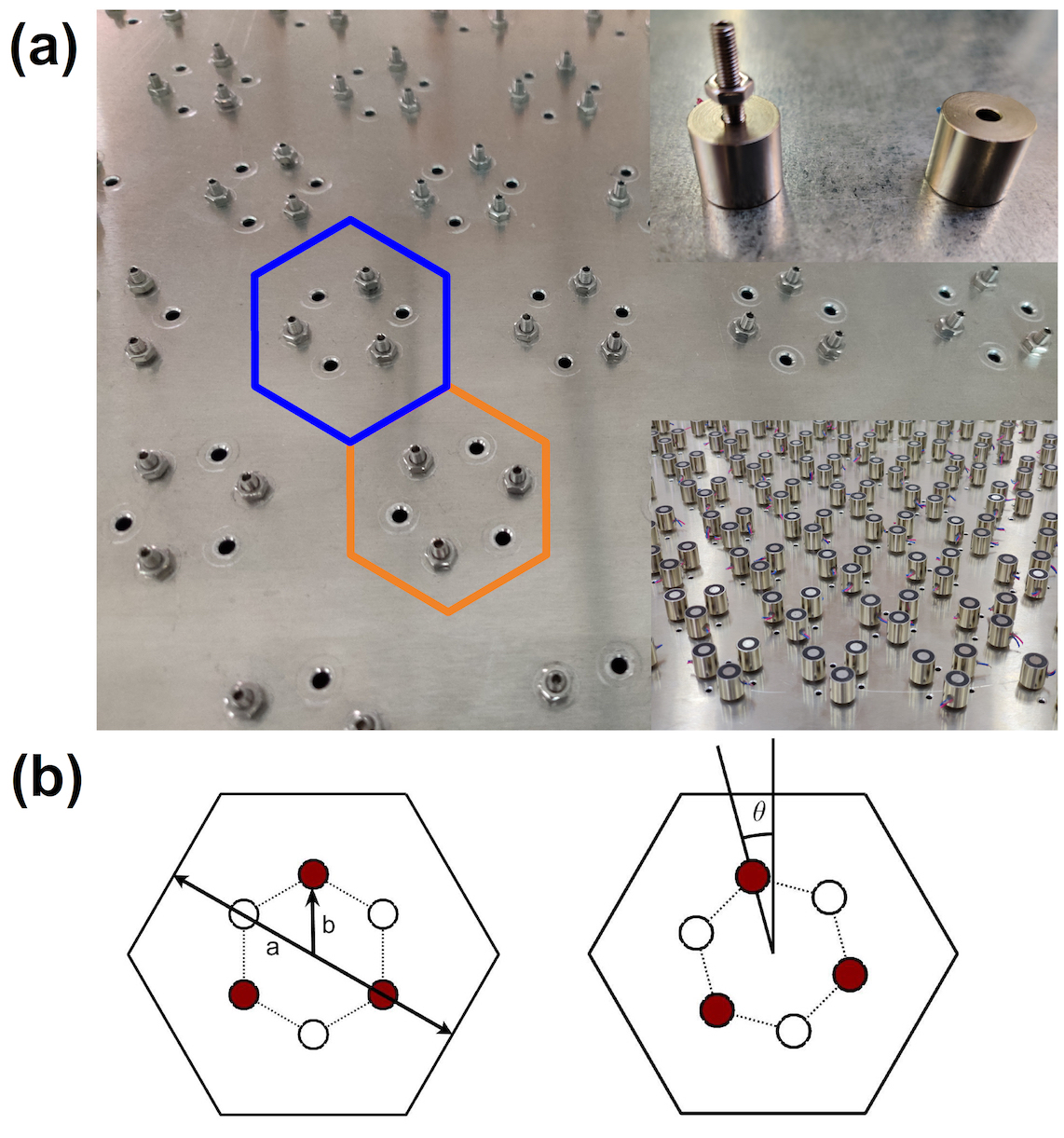} 
\caption{(a) Experimental system showing the macroscale hexagonal lattice comprised of elementary hexagonal cells (materialized in blue and orange) containing alternations of free and clamped holes arranged to lie at the vertices of a smaller hexagon. 
Left inset: cylindrical magnets, with and without clamping screws, magnetically attached to a honeycomb breadboard. Right inset: back view of the structured plate. (b) Schematics of the unit cell for unperturbed (left) and perturbed (right) periodic structures, with $a(=52$mm) the lattice constant and $b(=12$mm) the distance between the holes and center of the unit cell. The red dots and empty circles depict clamped and free holes, respectively.
 The rotation angle $\theta$ indicates the perturbation away, of the internal inclusion set, from the $\sigma_v$ symmetry axis.}
\label{fig:fig1}
\end{figure}

The system we study, Fig.~\ref{fig:fig1}, consists of a macroscale hexagonal lattice comprised of elementary hexagonal cells in a 1.6~mm-thick aluminium rectangular plate (830~mm x 780~mm); each cell contains an alternation of free and clamped holes arranged to lie at the vertices of a smaller hexagon. The free holes (diameter 3mm) are relatively weak scatterers (working wavelength 33mm), and allow for a reconfigurable experimental set-up, as they are left either free or we introduce clamping to the holes. The main source of scattering is by using clamps to enforce zero displacement of the hole. 
  The primary advantage of clamped holes,
  over other scatterers, e.g. resonators \cite{chuansali18a,zhu_design_2018,yan_on-chip_2018,jinkyu2019,pennec2019}, holes \cite{yu_18a},
 thin ligaments 
  \cite{liu_tunable_2018,vila_experi,miniaci2018,miniaci2019,mei2019,ltw2019}
 is both their ability to trap energy in zero-frequency \cite{gautier_landscape} 
 and also to induce strong scatter at higher frequencies \cite{lefebvre_unveiling_2017} and this enables the observation of modal patterns and conversions discussed in this Letter. Experimentally, clamped holes are implemented by screws, passing through the holes, that attach to a permanent magnet and thence effectively clamp these screws to the honeycomb breadboard (with optimized low-frequency damping) due to magnetic attraction, see Fig. 1(a) insets; the magnet diameter and height is 12~mm. %

\begin{figure}[h!]
\includegraphics[width = 8.5cm]{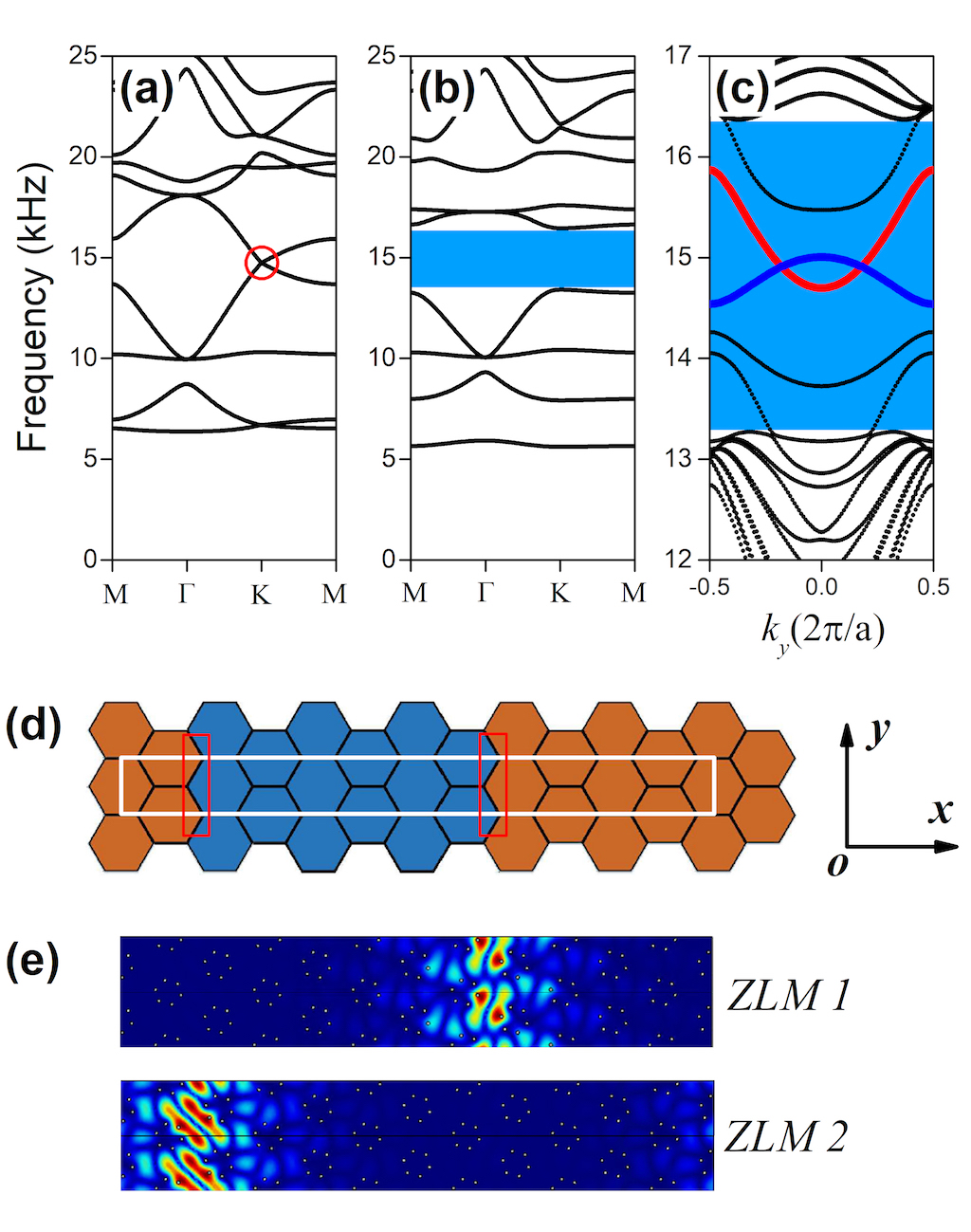} 
\centering
\caption{Band diagrams for (a) unperturbed, $\theta=0$, and (b) perturbed, $\theta=15^{o}$, periodic structures. In (a) the Dirac degeneracy is circled and in (b) the new band gap created by gapping the Dirac point is shaded light blue. (c) Dispersion curves with the ZLMs (red and blue curve) lying in the band bap shaded light blue. (d) Schematics of the finite ribbon (white rectangle) used to calculate the ZLMs (red rectangle) in physical space; the blue cell is rotated by $\pi/3$ from the orange cell. (e) Distinct mode shapes (two periods are shown here) are confined at the two different edges; the ZLMs shown here were computed within the white rectangle (panel (d)) near the frequency $14.80$kHz.}
\label{fig:band}
\end{figure}


Turning to the array itself, Fig.~{\ref{fig:fig1}(b)}, 
when the orientation angle of the hexagonal inclusion set is zero ($\theta$ = 0), the point group at $K$ (and $K^\prime$) has $C_{3v}$ symmetry and, provided we maintain the vertical mirror symmetry ($\sigma_v$), Dirac points are guaranteed to exist at the $K,K'$ points in the Brillouin zone (see Fig. 2(a)). Breaking the mirror symmetry, as we do by rotating with $\theta=15^ \circ$, removes  the $\sigma_v$ symmetry while retaining the $C_{3}$ symmetry and gaps the Dirac point to open a band gap, see Fig. 2(b), ranging from 13.4 kHz to 16.5 kHz. Numerically we consider the arrangement in Fig. 2(d) and  
 place the perturbed medium (blue) next to its $\pi/3$ rotated twin (orange) and this yields a band gap, common to both media, within which valley-Hall edge states, Fig. 2(c), are guaranteed to reside; 
 simulations are full 3D elasticity in 
  COMSOL  Multiphysics.
 We have created an  interface between two media having opposite valley-Chern numbers, at $K$ or $K'$, and these valley Hall edge states are aptly named zero-line modes (ZLMs) \cite{makwana_geometry_2018}. 
 Notably, both, the perturbed cell and its $\pi/3$ rotated twin can be constructed within the same plate without creating further holes. Hence, we are able to navigate energy in vastly different directions by clamping different holes within the same reconfigurable plate. Numerically we consider a finite ribbon (white rectangle in Fig.~{\ref{fig:band}(d)}), consider Bloch conditions in the $y$ direction, and apply periodic conditions at the left/right end of the ribbon; this modeling choice captures all the admissible edge states in a single model. The lack of reflectional symmetry in the perturbed structures, resulting in asymmetric edges, yields two distinct interfaces between two topologically distinct media (red rectangles in Fig.~{\ref{fig:band}(d)}). We therefore obtain two overlapping broadband ZLMs along these interfaces (red and blue curves in Fig.~{\ref{fig:band}(c)}). Both the modes for blue over orange (ZLM 1) and vice versa (ZLM 2) exist over a simultaneous frequency range (14.70 kHz to 15.00 kHz) and are distinguished in Fig.~{\ref{fig:band}(e)} via the different modal patterns. For valley-Hall states this is not the only way to generate edge states with hexagonal structures and a classification is available  \cite{makwana_geometry_2018} giving two additional  constructions for $C_{3v}$ interfaces not considered here.  
 



\begin{figure}[h!]
\centering 
\includegraphics[width=8.5cm]{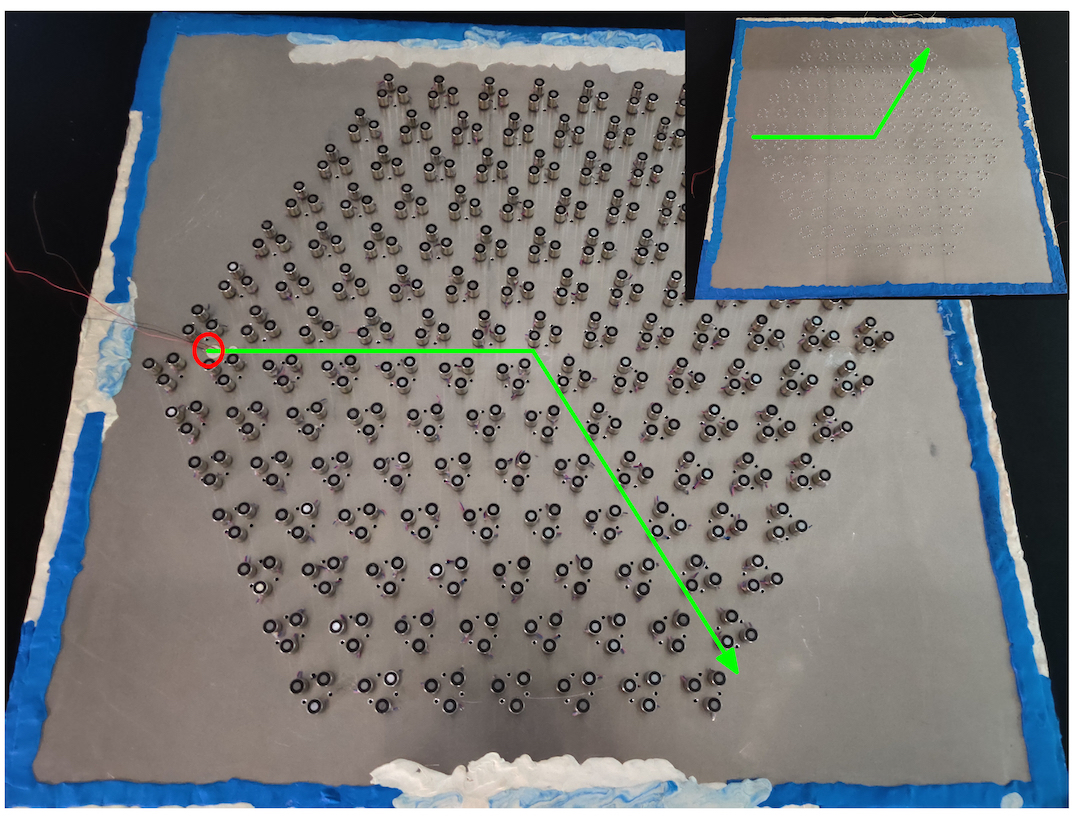} 
\caption{Experimental system: back view. Blu tack is attached to the edges of the plate to reduce unwanted elastic edge reflections; A piezoelectric transducer 
 is placed at leftmost edge (positioned with the red circle). The green arrows indicate the interface between two different media (here exemplified with gentle bending). Inset: Front view of the structured plate. The out-of-plane displacement at each point within the system is measured with a broadband heterodyne interferometric laser probe. 
 }
\label{fig:experi_setup}
\end{figure}

\begin{figure*} 
 \centering  
\includegraphics[width = 16cm]{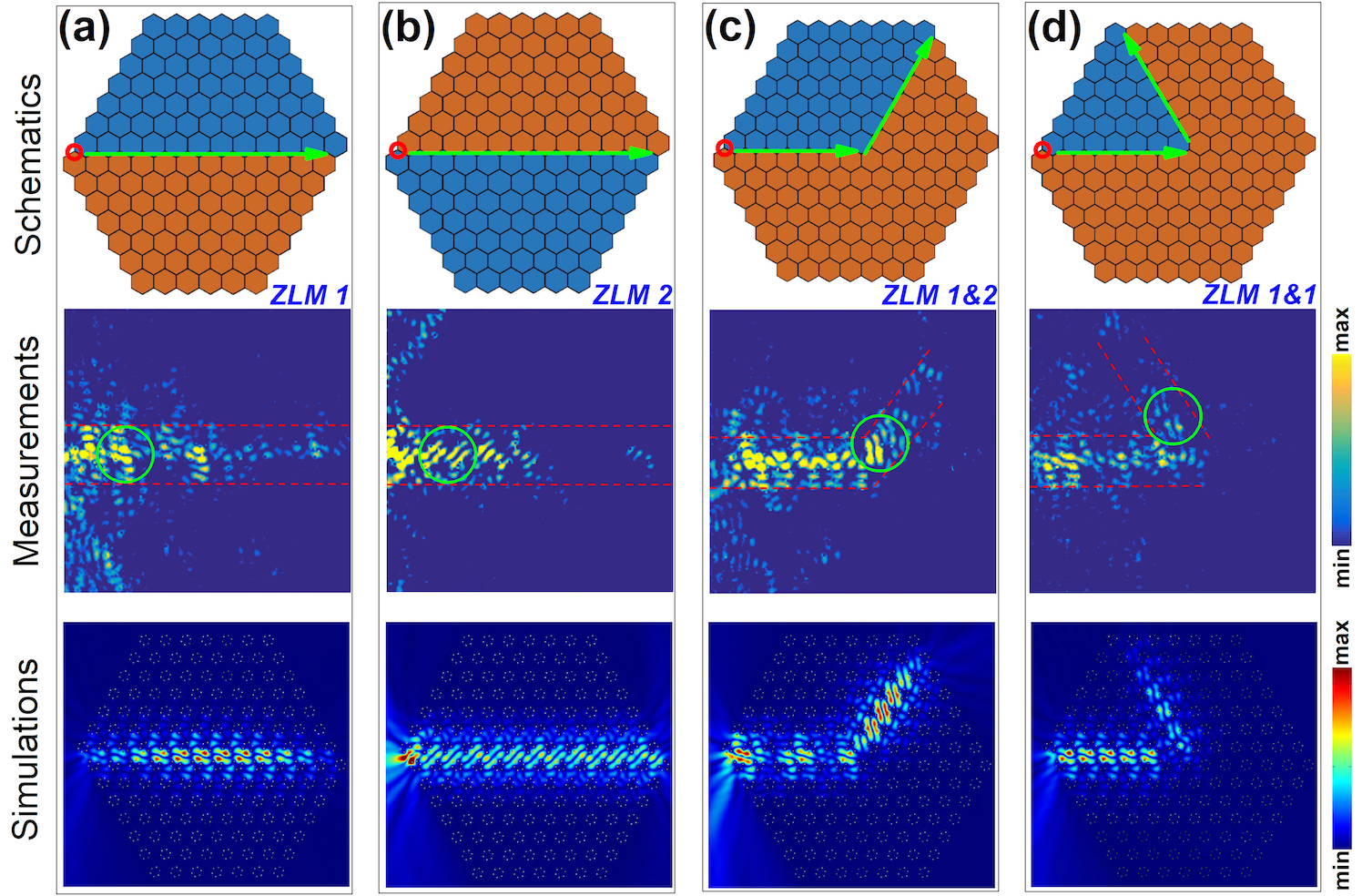} 
  \caption{First row, schematics for four different configurations: (a),(b) straight waveguides for ZLMs; (c) gentle bend, (d) sharp bend; red circles indicates source position and green arrows represent the anticipated directions of energy flux. Second row, Experimental measurements: Intensity distribution at (a)15.10 kHz, (b)15.30 kHz, (c)14.70 kHz, (d)15.12 kHz, respectively. The field patterns (highlighted with green circles), and intensities, are distinct for the two arms in (c), but identical for panel (d), which demonstrate the topological mode coupling between two overlapping ZLMs for the former and single ZLM mode transport for the latter.  
  The frequency of operation is chosen in the range 14.70 kHz and 15.00 kHz where both modes coexist (between the red and blue curves in Fig.~{\ref{fig:band}(c)}). This is essential for the gentle bend, to observe mode convesion from ZLM1 to ZLM2 at the bend. For an excitation frequency with no spectral overlap, then there is no coupling between the arms for gentle bend. Third row: Numerical simulations at frequency equal to (a)15.00 kHz, (b)15.30 kHz, (c)14.73 kHz, (d)15.00 kHz, respectively. The choice of excitation frequencies is elaborated on in the \emph{supplementary material} \cite{supp} and  \href{https://photos.app.goo.gl/wseevGNPyVfuJnpJ9}{\emph{movies} of our experimental results are also available}.}
  \label{fig:experi}
\end{figure*} 


The consequences of overlapping edge modes (in Fig.2(c)), for four different configurations (blue over orange and orange over blue ZLMs in addition to the gentle and sharp bends), are shown by experimental observations with a finite lattice of 840 circular holes forming the structured plate in Fig.~{\ref{fig:experi_setup}}. 
The configurations between two topologically distinct media we choose are in Fig. 4: straight waveguides for ZLMs, gentle bend 
(as in Fig. 3 where the green arrows indicate the anticipated direction of energy transport), and sharp bend,  
 and each is enabled by simply changing the positions of clamped holes using the magnets. The plate is excited using a broadband piezoelectric transducer (Murata 7BB-20-6) bonded to the back of the pinned plate, which functions as a quasi-point source and is positioned at the red circle in Fig. 3, that generates a Gaussian-modulated chirp pulse excitation, from 10 kHz to 20 kHz, synthesized by an arbitrary waveform generator (Agilent 33120A). The out-of-plane displacement is measured point-by-point with a broadband heterodyne interferometric laser probe (Polytec sensor head OFV534, controller OFV2500). The probe is scanned on a square grid with 4 mm (around one-eighth of the working wavelength) step resolution and the spatio-temporal distribution of the vibration field at the surface of the plate is reconstructed.  We employ Fast Fourier Transformations on the measured temporal signal and construct a 2D field distribution for a specific frequency. A Hampel filter function and a cubic interpolation  are applied to the measured grid data for a better visualization of wavefields on the plate. To prevent unwanted reflections, the plate edges are covered by a 2 mm-thick layer of blu-tack on both sides over 2 cm, which acts as a good absorber in the kHz range \cite{lefebvre_unveiling_2017}.

\begin{figure}[h!]
\centering 
 \includegraphics[width=8.5cm]{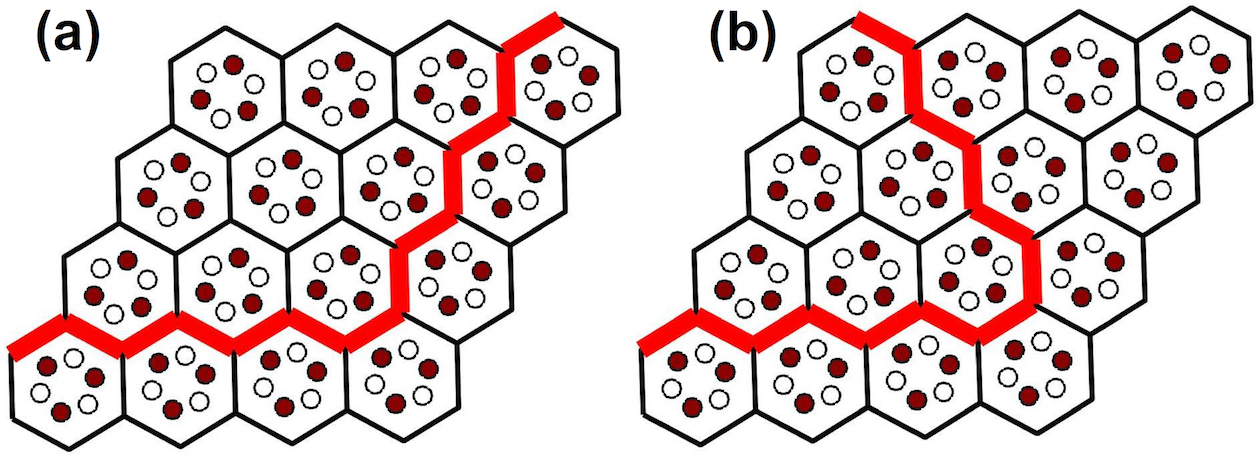} 
\caption{The detail of the junction cells, showing their asymmetric edges for gentle bending (a) and sharp bending (b).}
\label{fig:junction}
\end{figure}

The experimental results are in the second row of Fig.~{\ref{fig:experi}} where the red dashed lines show a width of two cells across the interface. Due to the scale of our experiment, we do not simply show the propagation path of the topological modes, but we also distinguish the detailed features of mode patterns due to the strong trapping of energy by the clamped holes. Initially, we demonstrate two straight waveguides, Fig.~{\ref{fig:experi}}(a) and (b);  
 the ZLM is excited at $f=15.10$ kHz and $15.30$ kHz, respectively 
 from the leftmost interface. Depending on the relative orientation of the interface (orange over blue or vice-versa), distinct  modal patterns (marked by green circles) are revealed similar to the simulations in Fig. 2(e).  
Then we analyze the gentle bend (Fig.~{\ref{fig:experi}(c)}), and again launch a ZLM from the leftmost interface towards the bend. Fig.~{\ref{fig:junction}(a)} shows a closeup of the four-by-four cells at the bend; from this figure it is difficult to ascertain the expected modal pattern along either of the interfaces along the gentle bend. However, the distinct modal patterns, pre- and post-bend are clearly evident in the experiments, Fig.~{\ref{fig:experi}(c)}; this indicates that there is strong coupling between one ZLM into the other (see Fig. 2(c)).

 Turning our attention to the sharp bend, and following the same thought process as above, one would naively expect the same behavior to that of the gentle bend. However, this is not the case as now the zigzag edges pre- and post-bend are identical; this is more easily seen from the scattering (Fig.~{\ref{fig:experi}(d)}) rather than the cellular arrangement (Fig.~{\ref{fig:junction}(b)}) and we see that  
 we now have identical modal patterns pre- and post-bend. The Z-shaped interface, comprising of two consecutive sharp bends, is often shown (\cite{zhang_topological_2018, zhang_manipulation_2018, yan_on-chip_2018, xiaoxiao_direct_2017, shalaev_experimental_2017, lu_observation_2016, liu_experimental_2018, jung_active_2018, gao_valley_2017, chen_tunable_2018, chen2019valley}) however the constituent modal patterns are often obscured. Here we show experimentally and with greater clarity the coupling between identical modes for this particular case.
 
 In practical terms, there is relatively low intensity in the second arm of Figs. 4(c) that is due to losses in our system. The losses are attributed to the out-of-plane coupling, between the plate and the air as well as the substrate (damping breadboard).
 This is not due to backscattering as  the robustness of the modes at the sharp bend is still topologically protected as seen in the corresponding simulations of the third row of Fig.~{\ref{fig:experi}}. The term ``robustness", related to topological modes, has two crucial meanings that are often conflated; robustness may pertain to either, the guaranteed existence of the valley-Hall states \cite{fefferman_edge_2015} (in the, presence or absence, of defects) or the reduction in backscattering of the edge states upon encountering a defect (for example, removing several inclusions close to the interface or encountering a junction that lies between several interfaces). Due to the impedance mismatch between the free plate at the input and output, the finite-size waveguides behave like a resonator cavity and impose discrete optimal working frequencies near the resonances. More details about the choices of working frequencies in the experiments are given in the supplementary materials. From the perspective of the simulations, the mode patterns are very clear and also present in the experiments although degraded due to losses that are not taken into consideration in the simulations. 
 
The importance of the different modes is that coupling around a bend is not guaranteed, in particular within an experimental context and for the gentle bend, as this requires overlap of the ZLMs. The wavefields in the second arm of Fig.~\ref{fig:experi}(c) are different from the first arm, which reveals there is mode coupling from ZLM 1 and ZLM 2, so long as both modes exist at the same excitation frequency; whereas the wavefields in both the first and second arms of sharp bending are identical (Fig.~{\ref{fig:experi}(d)}). If the ZLMs do not overlap at the operating frequency then there will be no topological mode coupling across the bend. According to the dispersion curves of Fig. 2(c), we observe mode coupling in simulations over the
frequency range from 14.70~kHz to 15.00~kHz, whereas we present here the measurements at frequencies which demonstrate the detailed mode patterns clearly (as highlighted by the green circles in Fig. 4). The small difference between experimental frequencies and numerical prediction is attributed to the absence of residual reflections in the simulations where perfectly absorbing boundaries allow full establishment of the modes. To further strengthen our arguments we have included additional experimental results in our supplementary material. These additional results are consistent with those shown in Fig. \ref{fig:experi}.

Experimental observations of topological valley transport around sharp  and gentle bends and the topological mode coupling around the gentle bend, for flexural waves highlight  differences between sharp and gentle bends and the importance of the relative orientations of inclusion sets on either side of the interface. The direct visualization of the mode patterns through spatial scanning of the wavefield, linked with the underlying principles at the junction cells, sheds light on the transport of energy around bends in partitioned media. We anticipate this insight will motivate the design of more efficient interfacial waveguides, topological networks \cite{makwana_design_2018}, and energy filters.

\begin{acknowledgments}
This research was supported in part by The Israel Science Foundation (Grants No. 1871/15 and 2074/15) and the United States-Israel Binational Science Foundation NSF/BSF (Grant No. 2015694). 
R.V.C. thanks the EPSRC (UK) for support and the Leverhulme trust for a Research fellowship.  
P. S. is thankful to the CNRS support under grant PICS-ALAMO.
\end{acknowledgments}

\bibliographystyle{apsrev}

\pagebreak
\widetext
\begin{center}
\textbf{{\myfont Supplementary Material}}
\end{center}
\setcounter{equation}{0}
\setcounter{figure}{0}
\setcounter{table}{0}
\setcounter{page}{1}
\makeatletter
\renewcommand{\theequation}{S\arabic{equation}}
\renewcommand{\thefigure}{S\arabic{figure}}
\renewcommand{\bibnumfmt}[1]{[S#1]}
\renewcommand{\citenumfont}[1]{S#1}

\section{Optimal frequencies for topological mode propagation}

\subsection{Frequency analysis for simulations}
We justify the choice of frequencies made in Fig.~4 by further examining the four configurations that lead to opposite parity ZLMs and propagation around gentle and sharp bends.
In the band diagram, Fig. 2(c), the blue and red branches correspond to the dispersion curves of the two edge states, ZLM1 and ZLM2, respectively. ZLM1 inhabits the frequency range  [14.54, 15.01] kHz, whilst ZLM2 modes belong to the frequency range [14.70, 15.87] kHz. For this band diagram we utilised the supercell method whereby we assumed that we had an infinite periodic system comprised of an array of ribbons (shown in Fig. 2(d)). Despite our focus being on the ZLMs that decay away from the interface this assumption still leads to discrepancies between the simulations and the physically bounded experimental system.

\begin{figure}[h!]
\includegraphics[width = 15cm]{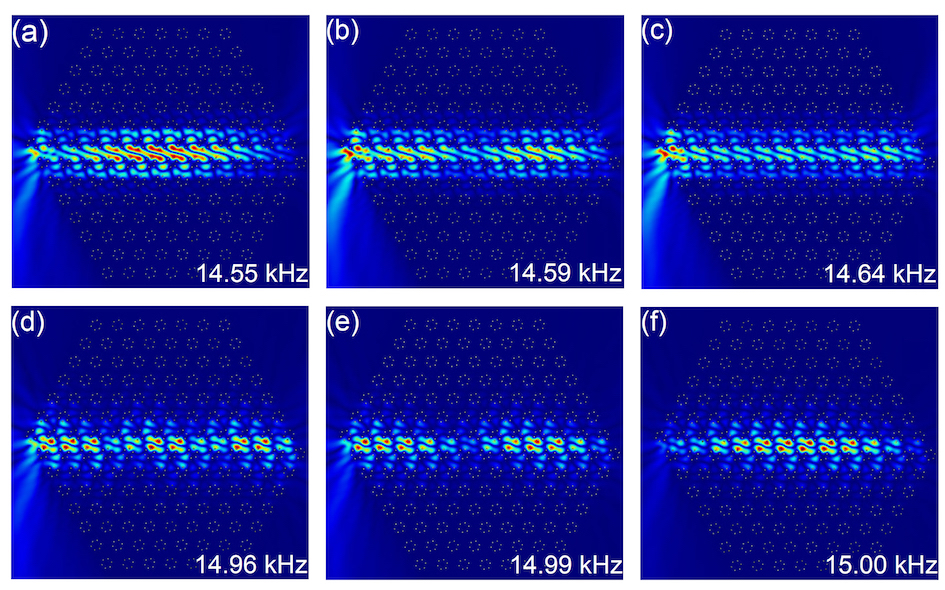} 
\caption{Scattering simulations for ZLM1 at resonant frequencies close to the standing wave. Displacements with 1, 2 and 3 antinodes are found near the upper (15.01kHz) and lower (14.54kHz) edges of the dispersion curve. The chosen resonant frequencies are shown as white circles in Figs. S4(a, b).}
\end{figure}

\begin{figure}[h!]
 \centering  
\includegraphics[width = 15cm]{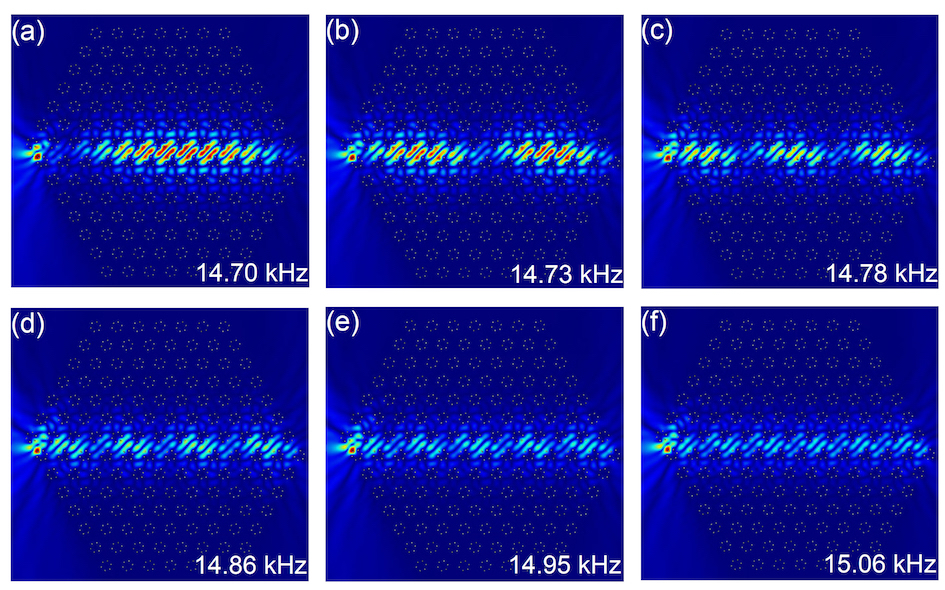} 
  \caption{Scattering simulations for ZLM2 at resonant frequencies close to the standing wave. Displacements that contain 1, 2, 3, 4, 5 and 6 antinodes are found starting from the lower edge (14.70kHz) of the dispersion curve. The chosen resonant frequencies are shown as white circles in Figs. S4(a, b).}
\end{figure}

Knowledge of the finiteness of our system is especially important for these passive topological modes as it can be used to minimise back-reflections. There is an impedance mismatch at the furthest end of our structured system than can result in enhanced backscattering. The ensuing superposition of forward and backwards travelling modes often leads to oscillations that have a discernible long-scale modulation. This behaviour is minimised when there are an integer number of wavelengths completely contained in each finite length interface. This resonance condition is formally written as, $k_y \times L = q \times \pi$ where $k_y$ is the wavevector along the waveguide, $L$ is the length of the waveguide and $q$ is an integer. This reduces our continuum set of solutions down to a discrete set of resonant solutions. We numerically study this behaviour near the standing wave solutions that delineate ZLM1's frequency range [14.5, 15.4] kHz in Fig. S1. Notably, a discrete set of antinodes appear when we are in the vicinity of either the upper cutoff frequency (15.01 kHz) or the lower cutoff frequency (14.54 kHz). A similar analysis is conducted for ZLM2 (Fig. S2) and for the gentle and sharp bends (Fig. S3).

\begin{figure}[h!] 
 \centering  
\includegraphics[width = 15cm]{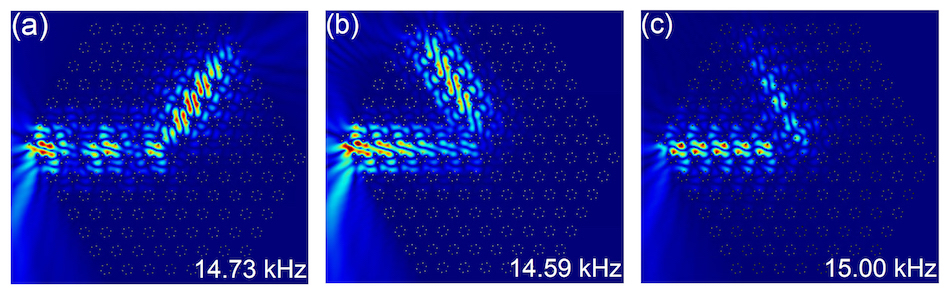} 
  \caption{Scattering simulations for (a) the gentle bend and (b),(c) the sharp bend at frequencies corresponding to post-bend resonance conditions. These frequencies are shown as white circles in Fig. S4(c).} 
\end{figure} 

In Fig. S4 we unearth the resonant frequencies by seeking out the maxima of the integrated amplitude for each of our designs. The original dispersion curves and the integrated amplitude variation for the ZLMs are shown in Figs. S4(a, b). Notably, the sharpest peaks in Fig. S4(b) (highlighted by the circles) corresponds to the band edge frequencies whose amplitudes were illustrated in Figs. S1, S2.

The integrated displacements for the gentle and sharp bends are shown in Fig. S4(c, d). Due to the shorter length of the straight interfaces for these designs the resonant condition differs from those of the ZLM: 
$k_y \times L/2=q \times \pi$. As our interest is in the coupling between the pre- and post-bend ZLMs, we opt to  examine the pre-bend and post-bend ZLMs integrated amplitudes separately. The resonant scattering solutions (highlighted circles in Fig. S4(c, d)) for these geometries are shown in Fig. S3. These correspond to a discrete frequency set in which the coupling between the incoming and outgoing ZLMs is optimal. Notably for the gentle bend, Fig. S4(c), the maximum value in the second half (at 14.73kHz) coincides with the minimum in the first half. This indicates that this specific frequency results in an optimal (almost perfect) transfer of energy of energy from the pre-bend mode (ZLM1) to the post-bend mode (ZLM2).

\subsection{Correspondence between simulations and experimentals}

As mentioned earlier, there is a natural distinction between our simulated dispersion curves (Fig. S4(a)) and the experiment. Not only is the finiteness of our system a factor, there are also out-of-plane losses to consider. Despite these factors, there remains a reasonable correspondence between the simulations and experiments whereby the former informs the latter. The optimal excitation frequency for the experimentally realised ZLM1 was found to be 15.1 kHz which is close to the sharpest simulated peak of 15.0 kHz (Fig. S4(b)). The sharp bend, which consists of an incident ZLM1 coupling to an outgoing ZLM1, has an optimal excitation frequency of 15.12 kHz close to ZLM1's resonant peak (Fig. S4(b)). Remarkably, the best experimental result for the gentle bend is found at 14.7 kHz which lies incredibly close to the simulated peak of 14.73kHz (Fig. S4(c)). This difference of 30 Hz may be further reduced by using a higher spectral resolution (our current experiment had a resolution of 20 Hz). Unfortunately, due to coupling issues between the source and the waveguide we were unable to experimentally demonstrate the propagation of ZLM2 for frequencies lower than 15.3 kHz.


\begin{figure}[h!]
 \centering  
\includegraphics[width = 15cm]{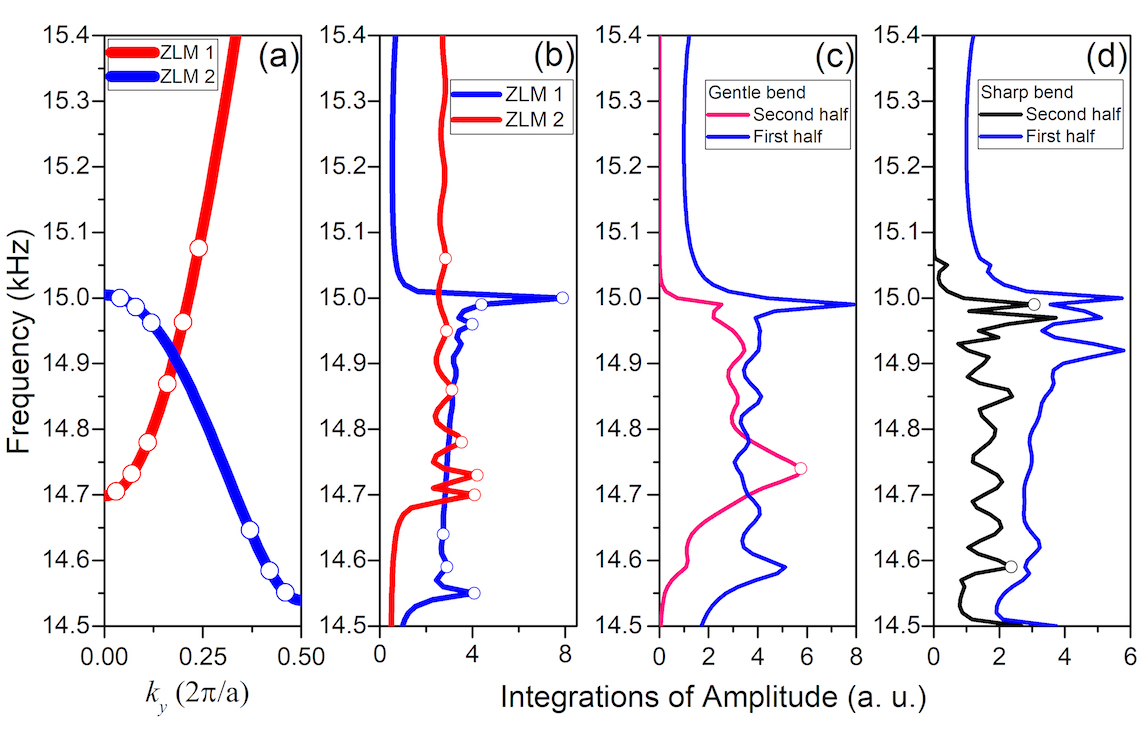} 
  \caption{(a) ZLM1 (blue) and ZLM2 (red) dispersion curves. The modal patterns exhibited in Figs. S1 and S2 are indicated by the white circles. (b) Integrated modal amplitudes vs. frequency for ZLM1 (blue) and ZLM2 (red). Integration was performed along a line along the straight waveguide, Fig. 4(a, b).  Integrated modal amplitudes vs. frequency for the (c) gentle bend and (d) the sharp bend. In this instance, integrations was performed along the pre-bend and  post-bend waveguide, respectively, Figs. 4(c, d). White circles in panels (b), (c) and (d) indicate resonant peaks that correspond to maximal transmission.}
\end{figure} 

\twocolumngrid

\end{document}